\documentclass[prb,aps,twocolumn,showpacs,floatfix]{revtex4-2}
\usepackage{amsfonts}
\usepackage{amsmath}
\usepackage{bbm}
\usepackage{hyperref}
\usepackage{graphicx}
\usepackage{dcolumn}
\usepackage{bm}
\usepackage{subfigure}
\usepackage{amssymb}
 \newcommand{\mi}{\mathrm{i}}\newcommand{\piup}{\mathrm{\pi}}
\newcommand{\dif}{\mathop{}\!\mathrm{d}}
\renewcommand{\vec}[1]{\bm{#1}}
\DeclareMathAlphabet{\mathfsl}{OT1}{cmss}{m}{sl}
\begin{document}
	\title{Current density in the anomalous Hall effect regime under weak scattering}
	\author{Ning Dai$^{1,2}$}
	\email{daining@hubu.edu.cn}
	\author{Bin Zhou$^{1,2}$}
	\affiliation{$^1$Department of Physics, Hubei University, Wuhan 430062, China;}
	\affiliation{$^2$Key Laboratory of Intelligent Sensing System and Security of Ministry of Education, Hubei University, Wuhan 430062, China.}
	\begin{abstract}
		A finite equilibrium current density arises in the anomalous Hall effect (AHE) as a result of time-reversal symmetry breaking, affecting both the differential current density and total current. This study illustrates the equilibrium current density pattern in a ribbon-shaped system within the AHE regime, consisting of two sets of counterpropagating channels arranged in a zebra crossing pattern. While the middle channels are susceptible to scattering, the edge channels remain relatively robust. Despite this difference, all channels exhibit the same differential current density when subjected to a differential voltage across the two ends of the ribbon. When a differential voltage is applied to both sides of the ribbon, it results in a snaking pattern of differential current density forming across it. Furthermore, in a four-terminal device comprising an AHE ribbon and two normal leads, it is found that Hall conductance is independent of ribbon width within certain scattering strengths due to differences in robustness between middle and edge channels. These findings underscore the significant role played by current density in AHE transport.
	\end{abstract}
\maketitle
\section{introduction}
The anomalous Hall effect (AHE) is the phenomenon of current carriers being deflected by the material's magnetization rather than an external magnetic field.\cite{1-1,1-2,1-3,1-4} This deflection arises from both extrinsic and intrinsic mechanisms, with disorder playing a key role in distinguishing between the two. Deflections induced by disorder are attributed to the extrinsic mechanism, which includes skew-scattering and side jump. Skew-scattering involves asymmetric scattering of current carriers on impurities,\cite{2-1} while side jump describes the displacement of the wave packet center at scattering sites due to disorder.\cite{2-2} On the other hand, the intrinsic mechanism explains that Berry curvature causes the wave packet to deviate from the external electric field.\cite{3-1,3-2,3-3}

Both extrinsic and intrinsic mechanisms explicitly demonstrate the AHE in a homogeneous material. In practice, mesoscopic transport in an AHE device inevitably involves the influence of the boundary. The role of the boundary is especially essential in the quantum AHE regime, where only boundary states remain due to the absence of bulk state.\cite{4-1,4-2,4-3,4-4,4-5,4-6} In general, both bulk and boundary current coexist in AHE, leading to a division of total current into bulk current and boundary current.\cite{5-1,5-2} The coexistence of bulk and boundary current exhibits novel phenomena, such as the antichiral edge state characterized by copropagating edge currents along with counterpropagating bulk current.\cite{6-1,6-2,6-3,6-4,6-5,6-6} Analysis on quantum transport in both boundary and bulk current is beneficial given this context. However, since the distinction between boundary and bulk currents is indistinguishable in the total current flowing into the terminals, it is necessary to conduct further investigation beyond the total current. Current density emerges as a promising candidate to uncover spatial details in AHE transport that are not captured by total current flow analysis.

While current density is widely utilized in mesoscopic transport research, it is often calculated as the component of differential current, defined as the increment of current following the addition of a differential voltage to an equilibrium state.\cite{7-1,7-2,7-3} This approach suffices for total current, as the equilibrium state always ensures zero total current through a contact. However, equilibrium current density is only guaranteed to be zero in time-reversal symmetric systems.\cite{8-1,8-2,8-3} A counterexample can be found in topological insulators (TIs), where edge states carry a current even without an external driving force.\cite{9-1,9-2,9-3,9-4,9-5} The equilibrium current density encompasses contributions from all Bloch waves below the Fermi surface, but the transport is primarily influenced by those near the Fermi surface. Time-reversal symmetry breaking in AHE preserves the possibility of a finite equilibrium current density.\cite{10-1,10-2} Unlike symmetry-protected edge currents in TIs, bulk equilibrium current density is fragile. Therefore, one important issue regarding equilibrium current density in AHE is its response to scattering, which has not been thoroughly studied yet.

In this paper, we use non-equilibrium Green's function to calculate both the equilibrium current density near the Fermi surface and the differential current density in a metallic time-reversal symmetry broken system. Such a system can be achieved by modulating the Fermi surface in a topological insulator to a metallic phase, hence referred to as topological metal (TM) below.\cite{11-1,11-2,11-3,11-4} B$\ddot{u}$ttiker probes (BPs) are utilized to model the inelastic scattering,\cite{12-1,12-2,12-3} as the intrinsic mechanism necessitates a voltage gradient across the device, which cannot be accomplished by elastic scattering processes. For an infinite TM ribbon, the equilibrium current density exhibits transverse oscillations, manifesting as two sets of counterpropagating channels arranged in a zebra crossing pattern.\cite{10-1} Adding a differential voltage to the two ends of the ribbon results in channels along the voltage drop experiencing an increase in flow, while those against the voltage drop experience depletion. Both sets of channels contribute the same differential current along the voltage drop. When scattering strength is gradually increased, channels in the middle of the ribbon are destroyed first, while channels near the boundary remain relatively robust. With a differential voltage added to both sides of the ribbon, a snaking pattern of differential current density forms across it. This pattern transforms into a uniform Hall current as scattering increases. Additionally, we examine a four-terminal AHE device by attaching two normal metal leads to both sides of the TM ribbon. Applying a differential voltage to these leads causes Hall current to flow along the TM ribbon. Due to its robustness compared to other channels, edge-channel conductance remains independent of ribbon width within certain scattering strengths. This example demonstrates that current density patterns in AHE have an impact on the total current.

\section{topological metal ribbon with longitudinal differential voltage}
\begin{figure}
	\includegraphics[width=\linewidth]{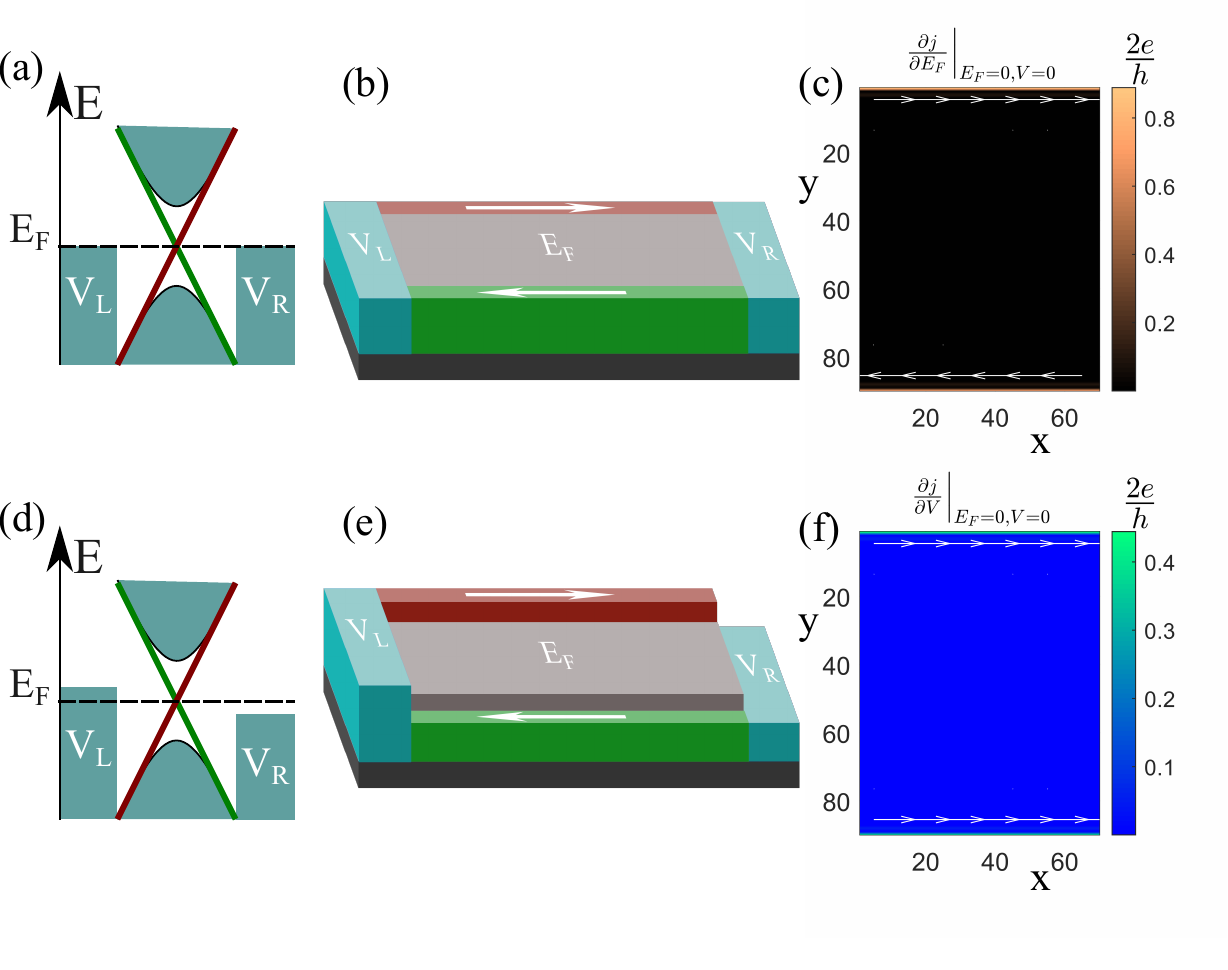}
	\caption{\label{fig1}Equilibrium (a-c) and non-equilibrium (d-f) transport in a topological insulator (TI) ribbon. (a,d) The potential distribution across the device. In the equilibrium state, the potentials in the two contacts are aligned with the Fermi surface, with only two edge modes crossing it. In the non-equilibrium situation, the differential voltage leads to a slight increase/decrease in potential at the left/right contact. (b,e) The schematic diagrams of ballistic transport in the TI ribbon. The two counterpropagating edge modes have an equal quantity of flow, resulting in a finite equilibrium current density and a zero total current. In the non-equilibrium situation, the mode towards the right/left gains/loses current corresponding to the differential voltage. (c,f) Numerical simulations of the equilibrium and differential current densities, with color representing the intensity and arrows indicating the direction of electronic flow. }
\end{figure}
We begin with a pedagogical quantum transport system to demonstrate the equilibrium and differential current density, which involves a TI ribbon connecting two contacts[see Fig.\ref{fig1}]. This system is in equilibrium when the voltages of the two contacts, $-V_L$ and $-V_R$, are equal, and can be described by a single Fermi energy $E_F$. In this paper, we use $-V_{L/R}$ to represent voltage, reserving $V_{L/R}$ for denoting the potential of electrons with negative charge. At zero temperature, the energy bands are completely filled below the Fermi energy and empty above it. The current density $\vec{j}$ is contributed by all the Bloch waves below the Fermi surface, but the response to the external fields is mainly contributed by those near the Fermi surface. In the case of TI, bulk states are far from the Fermi surface, leaving only two edge modes as two counterpropagating channels on both sides of the ribbon. These channels are filled with an equal amount of charged carriers, resulting in a finite equilibrium current density with zero total current. With a differential voltage $V=V_L-V_R$ applied, the chemical potential is slightly higher than $E_F$ in the left contact and slightly lower than $E_F$ in the right contact. The change in potential leads to two effects on transport. Firstly, incremental electrons from the left contact flow into the channel towards the right. Secondly, fewer electrons flow into the channel towards the left due to the decrease in potential in the right contact. The differential current density refers to the increment of the current density following the differential voltage, and both edge channels contribute to a rightward differential current density. Technically, we set the equilibrium Fermi energy $E_F=0$, and use $\left.\frac{\partial\vec{j}}{\partial E_F}\right|_{E_F=0,V=0}$ to represent the equilibrium current density near the Fermi surface, and $\left.\frac{\partial\vec{j}}{\partial V}\right|_{E_F=0,V=0}$ to represent the differential current density. Both current densities can be numerically simulated using the non-equilibrium Green's function method, and the details of the simulation are elaborated in the appendix.

\begin{figure}
	\includegraphics[width=\linewidth]{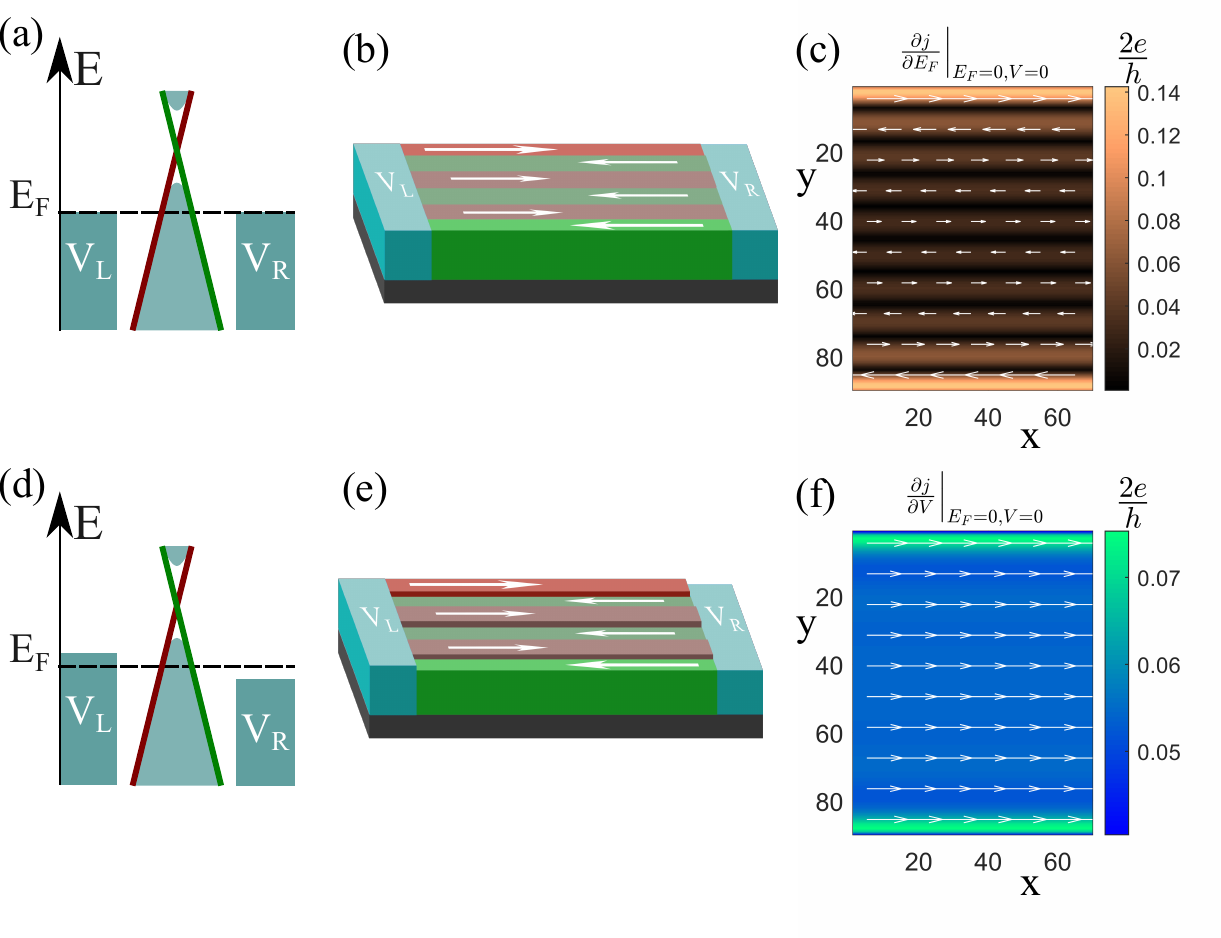}
	\caption{\label{fig2}Equilibrium (a-c) and non-equilibrium (d-f) transport in a topological metal (TM) ribbon. (a,d) The potential distribution across the device. The Fermi surface intersects the bulk states, transforming the TI into a TM. (b,e) The schematic diagrams of ballistic transport in the TM ribbon. A series of channels carry the current in alternating directions, which are distinguished by color in the diagram. The red channels propagate along the potential drop, while the green channels are in the reverse direction. The equality of quantities of flow in the red and green channels in equilibrium is disrupted by the differential voltage, generating a differential current density along the potential drop in all channels. (c,f) Numerical simulations of the equilibrium and differential current densities, with color representing the intensity and arrows indicating the direction of electronic flow.}
\end{figure}

By tuning the Fermi surface to intersect the bulk states, we transform the TI into a TM [Fig.\ref{fig2}(a)]. The TM ribbon lies on the x-y plane, with x along the length of the ribbon. The two contacts are located at the two ends of the ribbon and are assumed to be perfect, meaning that electrons in the TM ribbon can flow directly into the contact without undergoing reflection. In the equilibrium case, the current density near the Fermi Surface appears as a series of channels along the x direction. The current directions in these channels alternate along the y direction. When a differential voltage is applied, the forward channels witness an increase in flow, while the backward channels experience a decrease in flow. Compared with the equilibrium state, all channels contribute to a forward differential current density. 

The equilibrium current density oscillation stems from the combination of broken time-reversal symmetry and broken spatial translational symmetry. The bulk states in the ribbon comprise Bloch wave $|k_x>0\rangle$ and $|k_x<0\rangle$, carrying opposite current densities. Under time-reversal symmetry, each Kramers pair $|+k_x\rangle$ and $|-k_x\rangle$ have the same spatial distribution, resulting in a neutralization of current density. The breaking of time-reversal symmetry leads to a mass-center separation between $|+k_x\rangle$ and $|-k_x\rangle$, reserving a possibility for a non-trivial equilibrium current density distribution. However, if the translational symmetry along the y direction is maintained, the current density is uniform in y and can only be zero due to the zero total equilibrium current. Such symmetry can be examined in an infinite wide ribbon, or by applying periodical boundary condition to the ribbon to transform it into a tube. For this reason, the breaking of time-reversal symmetry generates a non-trivial equilibrium current density distribution only when the spatial translational symmetry is broken. The zebra crossing pattern of the equilibrium current density originates from the bulk and edge Bloch waves in the TM ribbon. Since all the bulk Bloch waves are simple harmonic in the y direction, all the middle channels have similar widths and flow intensities. This seemingly contradicts the thermodynamic limit, as a wide ribbon should approach an infinite plane. This contradiction results from the ideal perfect coherence model. In fact, the asymptotic behavior is illustrated under the influence of scattering.

\begin{figure}
	\includegraphics[width=\linewidth]{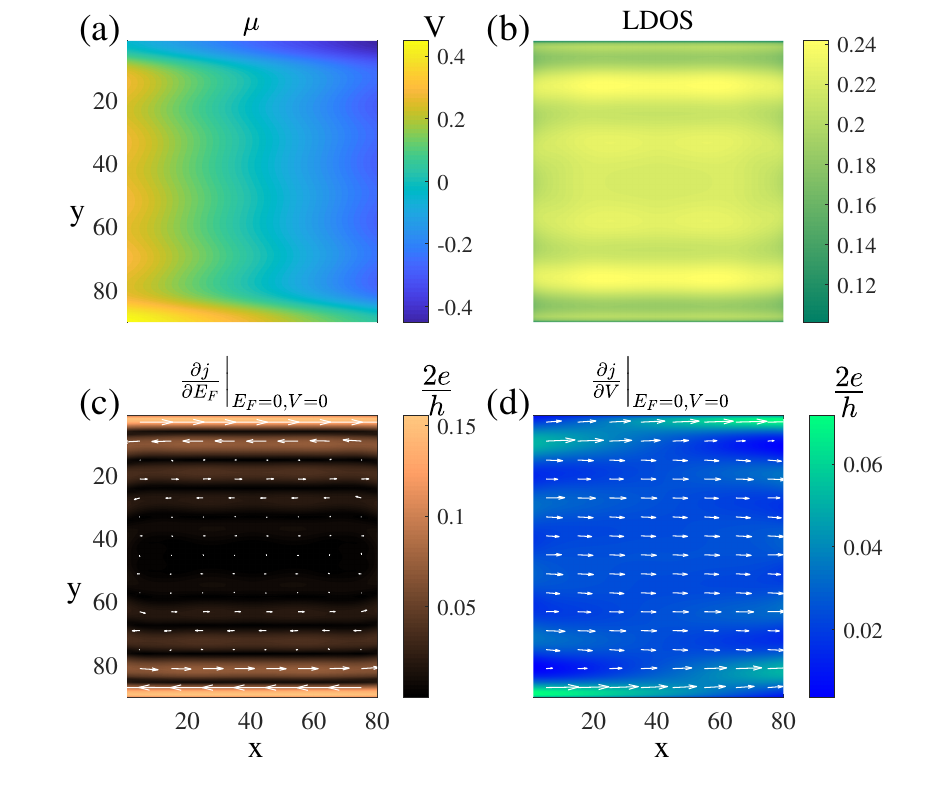}
	\caption{\label{fig3}Quantum transport in Fig.\ref{fig2} with the addition of weak scattering intensity $t_C=0.1$. (a) The chemical potential distribution across the TM ribbon under a differential voltage V. (b) Local density of state in the TM ribbon. (c) Equilibrium and (d) diffusive current density in the TM ribbon.}
\end{figure}

The scattering is simulated by B$\ddot{u}$ttiker probes (BPs). A BP is a fictitious contact with a zero total current, extracting electrons from the TM ribbon and reinjecting the same amount of electrons. The reinjected electrons lose all the information, including momentum, energy, and phase. Such reinjection can be regarded as an inelastic scattering at the terminal of the BP, and the coupling strength $t_C$ between BPs and the TM can be treated as the scattering intensity. Furthermore, the chemical potential $\mu$ in BPs reveals the chemical potential distribution across the TM ribbon [Fig.\ref{fig3}(a)]. In the presence of inelastic scattering and with a differential voltage $V$ applied, the chemical potential $\mu$ continuously decreases along the x direction from $\mu_L=0.5V$ in the left contact to $\mu_R=-0.5V$ in the right contact. However, this potential drop varies in the y direction. The most remarkable difference exists between the two edge modes at $y=0$ and $y=L_y$, where $L_y$ is the width of the ribbon. Electrons in the left contact flow into the $y=Ly$ edge channel without reflection and propagate under relatively low resistance, keeping $\mu$ almost in alignment with the left contact until the electrons are strongly deflected at the right contact. Due to the same mechanism, the opposite edge channel is rather in alignment with the right contact in terms of chemical potential. This mechanism also explains the potential drop in the middle channels, which occurs near the left contact in the leftward channels and near the right contact in the rightward channels. The chemical potential distribution influences the differential current density [Fig.\ref{fig3}(d)]. As the differential current density represents the increment between non-equilibrium and equilibrium, it is intense where $\mu$ is far from 0 and weak where $\mu$ is close to 0. In middle channels, the chemical potential reaches 0 earlier than in edge channels, indicating the different robustness of the channels. The impact of the scattering on channels increases with the channel's distance from the boundary [Fig.\ref{fig3}(c)], which explains the asymptotic behavior from a wide TM ribbon with finite equilibrium current density to an infinite TM plane without equilibrium current density. In the presence of the scattering, the equilibrium current density is only distributed near the boundary of a wide TM ribbon, and the major center region is current-free, resembling an infinite TM plane. It is worth noting that the difference in the robustness of channels is only manifested in the current density but indistinguishable in the local density of state (LDOS) [Fig.\ref{fig3}(b)].

\section{topological metal ribbon with transversal differential voltage}
\begin{figure}
	\includegraphics[width=\linewidth]{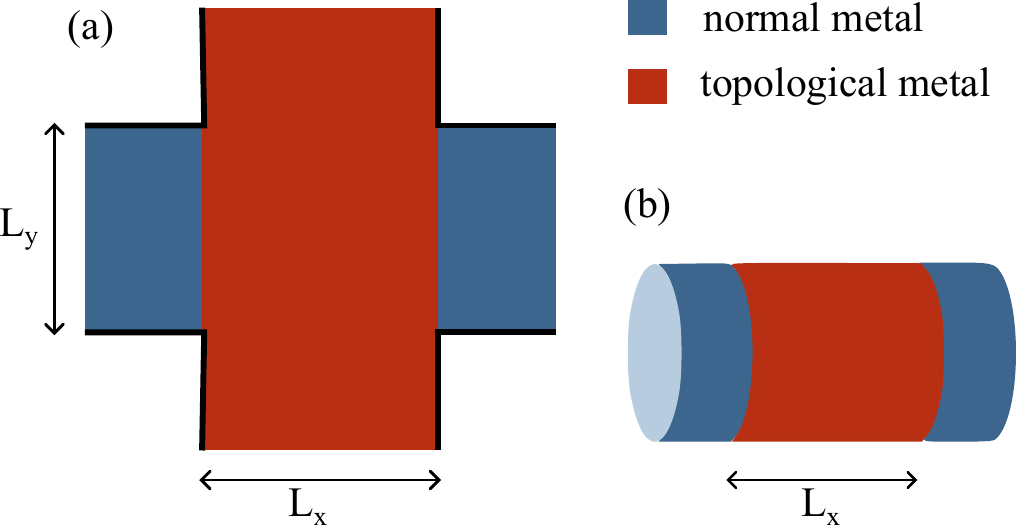}
	\caption{\label{fig4}(a) Four-terminal transport device composed of a TM ribbon and two normal metal contacts. A differential voltage $V$ is applied at the two contacts, generating a differential Hall current along the TM ribbon. (b) The tube-shaped two-terminal transport device, imitating the $Ly\to\infty$ limit in (a).}
\end{figure}

The AHE describes the relationship between the current and the corresponding perpendicular voltage, which is not manifested in the two-terminal device. Figure.\ref{fig4}(a) presents a typical four-terminal device for measuring the Hall conductance, consisting of a TM ribbon with two normal metal contacts. The TM ribbon extends along the y direction with the width $L_x$, and the two contacts attach to the ribbon symmetrically with the width $L_y$. A differential voltage $V$ is applied at the two contacts, generating a differential Hall current along the TM ribbon. However, although the crossing device resembles the real experiment, the deflection at the corner of the contact complicates the current density in the middle of the ribbon. To simulate a crossing device with wide contact, we apply the period boundary condition in the y direction and transform the crossing into a tube in Fig.\ref{fig4}.(b).

\begin{figure}
	\includegraphics[width=\linewidth]{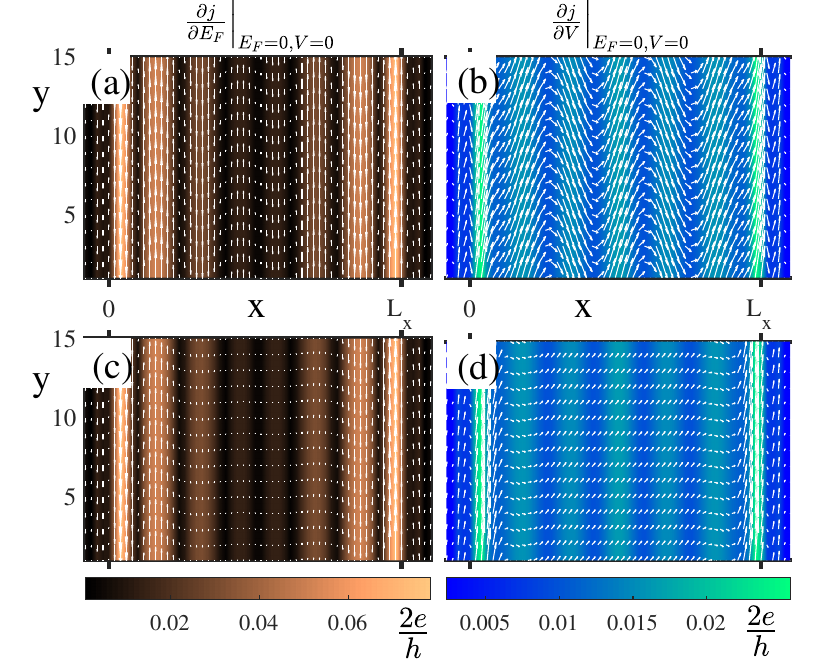}
	\caption{\label{fig5}(a,c) Equilibrium and (b,d) differential current density in the TM region of Fig.\ref{fig4}(b), with scattering intensity $t_C=0.1$ in (a,b) and $t_C=0.3$ in (c,d). The potential drop is along the x axis, while the y axis is along the circumferential direction of the tube.}
\end{figure}

The tube-shaped two-terminal system eliminates the deflection at the corner and serves as an excellent sample for investigating the $y\sim L_y/2$ region in the crossing device shown in Fig.\ref{fig4}.(a) with wide contacts. Figure.\ref{fig5} presents the equilibrium and differential current density in the tube-shaped system under $t_C=0.1$ and $t_C=0.3$ respectively. The channels extend along the y direction, corresponding to the circumferential direction of the tube. Due to the influence of scattering, the middle channels at $x\sim L_x/2$ are severely impaired, leading to a significant reduction in equilibrium current density. Meanwhile, the side channels near the two interfaces are relatively robust, maintaining the equilibrium current density at a certain level. Under weak scattering ($t_C=0.1$), the differential current is driven by the potential drop along the x direction, meandering through a series of alternating channels, resulting in a snake-like pattern of the differential current density. Under relatively strong scattering ($t_C=0.3$), the middle channels are significantly disrupted, transforming the snaking pattern of the differential current into a uniform Hall current in the middle.

\begin{figure}
	\includegraphics[width=\linewidth]{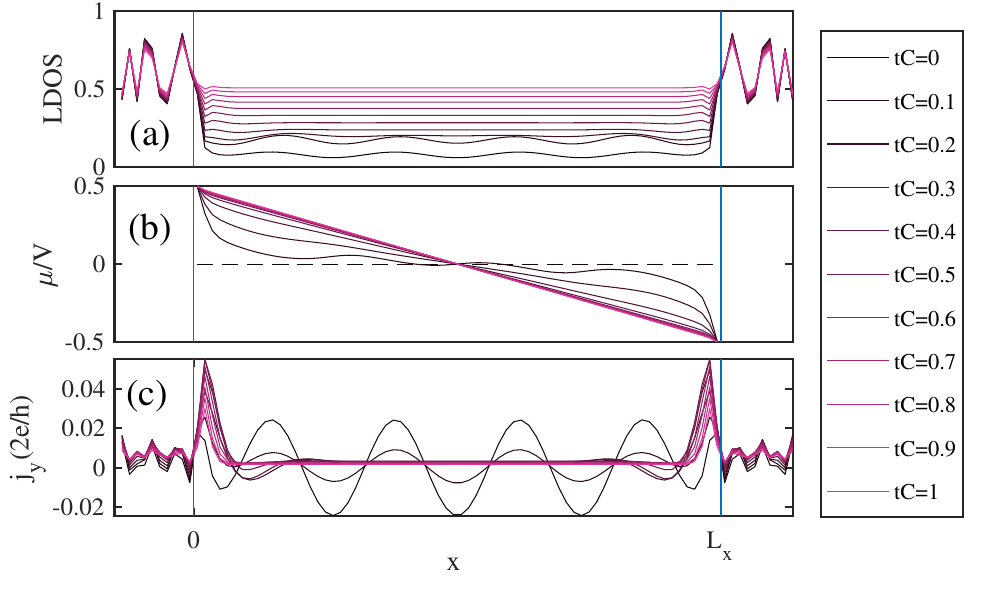}
	\caption{\label{fig6}The distribution of (a) LDOS, (b) chemical potential, and (c) Hall current density in the tube-shaped two terminal system portrayed in Fig.\ref{fig4}(b) under different intensities of scattering. The length of the TM region is $L_x=90$.}
\end{figure}

Since the tube is translational invariant in the y direction, we present the LDOS, chemical potential, and Hall current density versus the x coordinate under different scattering intensities in Fig.\ref{fig6}. In the pristine system without scattering, the LDOS is high in the wide normal metal contact and drops significantly in the TM near the interface. The LDOS in the TM increases with the scattering intensity and trends from oscillation to uniformity in this process, indicating the dephasing process. In the Ohmic transport regime with strong scattering, the chemical potential $\mu$ linearly decreases from 0.5V to -0.5V. While in the ballistic transport regime with weak scattering, the chemical potential drops dramatically near the terminal and oscillates in the middle region under the phase coherence effect. The most notable ingredient is the Hall current density $j_y$, which represents the y component of the differential current density  $\left.\frac{\partial\vec{j}}{\partial V}\right|_{E_F=0,V=0}$. By comparing the equilibrium and differential current density in Fig.\ref{fig5}(a) and (b), the disruption of the weak scattering $t_C=0.1$ behaves differently in the equilibrium and differential current density. While the middle channels suffer more damage than the side channels in the equilibrium current density, there is no distinct difference between the middle and side channels (expect the two outermost edge channels) in the differential current density. Such a difference is observable under a stronger scattering $t_C=0.3$ [see Fig.\ref{fig5}(d)]. In the weak scattering regime, the Hall current density oscillates uniformly over the majority of the x axis, and the amplitude decreases with the increase of the scattering intensity $t_C$. The differential current density in the weak scattering regime can be explained as follows: The voltage drop is perpendicular to the channels, such that the channels are not fully occupied by the differential current. Under weak scattering, the various degree of slight disruptions to the channels do not determine the differential current flux passing through the channel. Under relatively strong scattering when middle channels are severely disrupted, the differential current oscillation initially fade away in the center of the tube.

\begin{figure}
	\includegraphics[width=\linewidth]{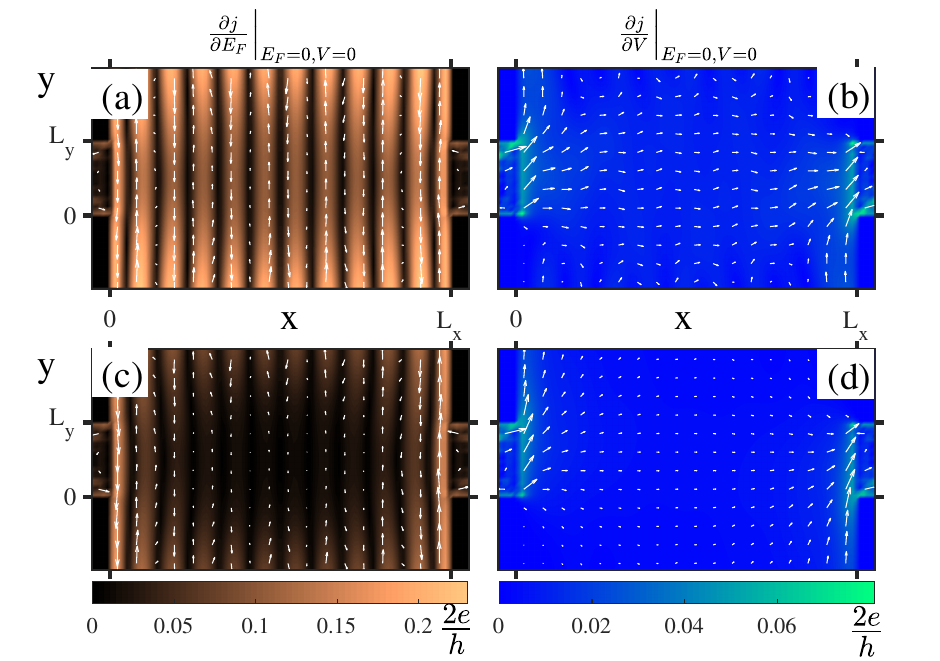}
	\caption{\label{fig7}(a,c) Equilibrium and (b,d) differential current density in the four-terminal system in Fig.\ref{fig4}(a), with scattering intensity $t_C=0$ in (a,b) and $t_C=0.1$ in (c,d). The width of the TM ribbon and the contacts are $L_x=90$ and $L_y=20$ respectively. The size of the extracted section is 100*60, with the abundance of 5 in each contact and 20 in each end of the ribbon.}
\end{figure}

Figure.\ref{fig7} illustrates the four-terminal system in Fig.\ref{fig4}(a) where the width of the TM ribbon and the contacts are $L_x=90$ and $L_y=20$ respectively. A section in the TM ribbon near the contacts is extracted, and numerical simulation is applied to obtain the equilibrium and diffusive current density. In the pristine system with $t_C=0$, the contacts slightly reduce the equilibrium channels in the central region, indicating the disruption of the translational symmetry in the y direction. When a differential voltage is applied, the snaking pattern emerges in the differential current density in the middle of the TM ribbon. In the presence of a finite scattering intensity $t_C=0.1$, both the equilibrium and differential current are significantly suppressed except in the two edge channels. Notably, the weak differential current density in the middle of the four-terminal system contrasts with the robust middle current in the tube-shaped system despite the same scattering intensity $t_C=0.1$ [see Fig.\ref{fig5}(a)]. This contrast stems from the different qualities of the contacts in Fig.\ref{fig4}(a) and (b), as the contacts in the tube are wider than in the crossing, injecting more electrons into the TM region.

\begin{figure}
	\includegraphics[width=\linewidth]{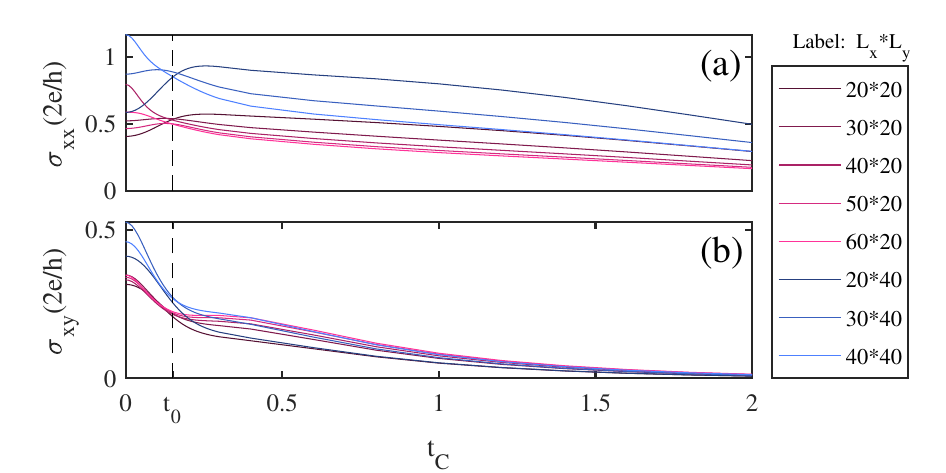}
	\caption{\label{fig8}(a) The longitudinal conductance $\sigma_{xx}$ and (b) transversal conductance $\sigma_{xy}$ varying with the scattering intensity $t_C$ in four-terminal samples of different sizes. The conductance with the narrow conducts $L_y=20$ is depicted by red curves, while that with the wide conducts $L_y=40$ is depicted by blue ones, and the color saturation indicates the width of the TM ribbon. The abundance in each contact and each end of the ribbon is the same as in Fig.\ref{fig7}.}
\end{figure}

We investigate both the longitudinal and transversal conductance varying with scattering in different-sized four-terminal samples, as shown in Fig.\ref{fig8}. The samples are divided into two groups based on the narrow ($L_y=20$) and wide ($L_y=40$) contact, and each group contains samples with different width of the TM ribbon $L_x$.  In the pristine system without scattering, the incidence at the interface is affected by the entire shape of the device due to the phase coherence effect, resulting in different longitudinal and transversal conductance at $t_C=0$. As $t_C$ increases, the Hall conductances of each group rapidly converge within a certain section, and then diverge near a critical scattering intensity $t_0=0.15$. Meanwhile, the longitudinal conductances in each group also briefly converge at $t_0$. The convergence section in Hall conductance indicates that the major capacity of Hall current is held by the two edge channels in the weak scattering regime with $t_C<t_0$. In this situation, the Hall conduction mainly depends on the flux in the edge channels, which is irrelevant to the width of the ribbon. When $t_C>t_0$, the weakened edge channels lose their domination over the bulk current, so the Hall conductance is related to the width of the bulk. The brief convergence in longitudinal conductance results from two different mechanisms with distinct effects. In the weak scattering regime, the dephasing effect reduces the variation of the injection at the interface, causing the longitudinal conductances to get closer as $t_C$ increases. When the scattering is intense enough to break the phase coherence, a stronger scattering forces electrons to follow a more tortuous propagation path, highlighting the difference in the shape of the device. The critical scattering intensity $t_0$ indicates the minimal variance generated by these two mechanisms. In the strong scattering regime, the longitudinal conductance is entirely dependent on the ratio $L_y:L_x$, which resembles the Ohm's law in a two-terminal transport system. On the other hand, the Hall conductance is mainly determined by the width of the TM ribbon $L_x$, because a wider TM ribbon offers more opportunities for electrons to deviate from the contacts and reach the terminal of the TM ribbon. The width of the contacts $L_y$ is rather insignificant to the Hall conductance, as only electrons near the boundary of the contacts have the possibility to deviate from the contacts.

\section{conclusion}
In summary, we have investigated the quantum transport in the AHE regime within a finite-sized device by illustrating the equilibrium and differential current density in a TM ribbon under various conditions. The equilibrium current density appears as a series of alternating channels along the ribbon, resembling a zebra crossing pattern. In the ballistic transport limit, all the middle channels have similar widths and flow intensities. However, under the influence of scattering disturbance, the middle channels collapse more rapidly than the side channels. With a differential voltage applied at both ends of the TM ribbon, all these channels contribute to the current density along the potential drop. The direction and robustness of the channel affect the distribution of the chemical potential, which determines the intensity of the differential current density. The AHE is examined in a four-terminal device consisting of a TM ribbon with two attached contacts. In the wide contact limit, the differential current meanders across the TM ribbon along a snaking path under weak scattering, and reduces to uniform Hall current density when scattering becomes more intense. Since the transverse channels are not fully occupied by the differential current, various degrees of slight disruptions to the channels do not determine the differential current flux passing through the channel. When the width of the contacts is finite, the longitudinal and Hall conductance are related to the width of the contacts as well as the TM ribbon. A critical scattering intensity denotes the minimal variance of the longitudinal conductance within the same width of the contact, and the Hall conductance is independent of the width of the ribbon within a certain scattering intensity section below the critical scattering intensity.

\begin{acknowledgments}
	We thank H.-L. Li for helpful discussion. N.D. was supported by the National Natural Science Foundation of China (Grant No. 12304062). B.Z. was supported by the NSFC (Grant No. 12074107), the program of outstanding young and middle-aged scientific and technological innovation teams of colleges and universities in Hubei Province (Grant No. T2020001), and the innovation group project of the Natural Science Foundation of Hubei Province of China (Grant No. 2022CFA012).
\end{acknowledgments}

\begin{appendix}
\section{model Hamiltonian}
In this Appendix, we introduce the model Hamiltonian in the numerical simulation, including the TI in Fig.\ref{fig1}, normal metal contacts in Fig.\ref{fig4}-\ref{fig8}, and the TM throughout the entire work. 

The two-band Hamiltonian of the TI is
\begin{equation}
	H_{TI}=\vec{d}(\vec{k})\cdot\vec{\sigma},
\end{equation}
with
\begin{eqnarray}
	d_x&=&A\sin k_xa,\\
	d_y&=&A\sin k_ya,\\
	d_z&=&\Delta-4B\sin^2\frac{k_xa}{2}-4B\sin^2\frac{k_ya}{2}.
\end{eqnarray}
The energy gap closes and reopens at three points: (1) $\Delta=0$; (2) $\Delta=4B$; (3) $\Delta=8B$, and topological quantum phase transition happens at all these points. In the case of $A>0$, the Chern number of the valence band is $C=0$ in $\Delta<0$ and $\Delta>8B$, $C=1$ in $0<\Delta<4B$, and $C=-1$ in $4B<\Delta<8B$. 

The Hamiltonian of the TM is
\begin{equation}
	H_{TM}=H_{TI}+\varepsilon\sigma_0,
\end{equation}
where $\varepsilon$ denotes the shift of the Fermi surface.

The natural units are adopted in this paper and the energy unit is chosen as $A=1$. The parameters of TMs in this paper are $B=1$, $\Delta=0.1$, and $\varepsilon=0.2$. Particularly, the TI in Fig.\ref{fig1} is obtained by removing the $\varepsilon\sigma_0$ term.

The numerical simulation is performed on the lattice model. The discretized lattice Hamiltonian can be expressed as the on-site Hamiltonian of a single lattice point $H_{00}^{(T)}$ and hopping terms between two adjacent lattice points $H_{0x}^{(T)}$ and $H_{0y}^{(T)}$.
\begin{eqnarray}
	H_{0x}^{(T)}&=&\frac{\mi}{2}A\sigma_x+B\sigma_z,\\
	H_{0y}^{(T)}&=&\frac{\mi}{2}A\sigma_y+B\sigma_z,\\
	H_{00}^{(T)}&=&(m-4B)\sigma_z+\mu\sigma_0.
\end{eqnarray}
Here $H_{0x}^{(T)}$ denotes the hopping term from coordinate $(x,y)$ to $(x+a,y)$, and $H_{0y}^{(T)}$ denotes the hopping term from coordinate $(x,y)$ to $(x,y+a)$, where $a$ is the lattice constant and is the unit length $a=1$ in the main text.

In addition, the discretized lattice Hamiltonian of the normal metal contact is
\begin{eqnarray}
	H_{0x}^{(N)}=H_{0y}^{(N)}=t_N\sigma_z,\\
	H_{00}^{(N)}=\varepsilon_N\sigma_z,
\end{eqnarray}
with $t_N=3$ and $\varepsilon_N=-2$.

\section{chemical potential distribution}
In the present and subsequent Appendix, we develop the Non-equilibrium Green's function method with B$\ddot{u}$ttiker probes to simulate the chemical potential distribution and the current density. Consider a lattice model connected to several real leads and many fictitious leads. The current flow in the $i$th lead is
\begin{equation}
	I_i(E)=\sum_{k\in\mathrm{all}}T_{ik}(E)[f_i(E)-f_k(E)],
\end{equation}
where the index $k$ runs over all real and fictitious leads. The scattering coefficient is
\begin{equation}
	T_{ik}(E)=\mathrm{Tr}[\Gamma_i(E)G^R(E)\Gamma_k(E)G^A(E)],
\end{equation}
where $G$ is the Green's function and $\Gamma_i$ is the linewidth function for the $i$th lead. The Fermi distribution function in zero temperature is
\begin{equation}
	f_i(E)=\theta(\mu_i-E).
\end{equation}
Here $\mu_i$ denotes the chemical potential in the $i$th lead. 

The total current in the lead
\begin{equation}
	I_i=\int I_i(E)\dif E
\end{equation}
is zero in the equilibrium state. When a differential voltage is applied, the total current is equal to the differential current
\begin{equation}
	I_i=\sum_{k\in\mathrm{all}}T_{ik}(\mu_i-\mu_k)
\end{equation}
However, if the $i$th lead is fictitious, the total current must be zero, which means
\begin{equation}
	I_i=\sum_{j\in\mathrm{fic}}T_{ij}(\mu_i-\mu_j)+\sum_{m\in\mathrm{real}}T_{im}(\mu_i-\mu_m)=0.
\end{equation}
We put this equation in the following form:
\begin{equation}
	\sum_{j\in\mathrm{fic}}T_{ij}(V_i-V_j)=\sum_{m\in\mathrm{real}}T_{im}(\mu_m-\mu_i).\label{1}
\end{equation}
Here the index $j$ runs over all fictitious leads, and $m$ runs over all real leads. Introduce the matrices $T$, $T_{diff}$ and vectors $\vec{\mu}$, $\vec{I}_{inci}$. Here $T_{ij}$ is the scattering coefficient within fictitious leads. $T_{diff}$ is a diagonal matrix where the $j$th element on the diagonal is $\sum_{k\in\mathrm{all}}T_{jk}$. The vector $\vec{\mu}$ consists of the chemical potentials in all the fictitious leads, and the $j$th element of $\vec{I}_{inci}$ is $\sum_{m\in\mathrm{real}}T_{jm}\mu_m$. With the help of these notation, Eq.\ref{1} is expressed in a matrix form:
\begin{equation}
	(T_\mathrm{diff}-T)\vec{\mu}=\vec{I}_{inci}.\label{2}
\end{equation}
Since the chemical potentials in the real leads are given, and the scattering coefficient can be obtained by the well developed Non-equilibrium Green's function method, the chemical potential distribution in fictitious leads $\vec{\mu}$ can thus be obtained through Eq.\ref{2}. In this paper, each lattice point in TM is connected to a fictitious lead with a bandwidth $t=1$, so $\vec{\mu}$ disclose the chemical potential distribution in the TM region.

\section{current density}
The current from site $\vec{r}$ to the adjacent site $\vec{r}'$ is
\begin{equation}
	I_{rr'}=-\frac{2e}{\hbar}\int\frac{d E}{2\piup}\sum_{i,j}\mathrm{Re}([t_{rr'}]_{ij}^*[G_{rr'}^<(E)]_{ij}),
\end{equation}
where $i,j$ run over all the orbitals. Here $t_{rr'}$ denotes the hopping matrix between $\vec{r}$ and $\vec{r}'$ in the discretized lattice Hamiltonian.

In the equilibrium state, the lesser Green's function can be obtained by the fluctuation-dissipation theorem,
\begin{equation}
	G^<(E)=-f(E)[G^r(E)-G^a(E)].
\end{equation}
The Fermi distribution function in zero temperature is
\begin{equation}
	f_i(E)=\theta(E_F-E),
\end{equation}
where $E_F$ is the Fermi energy of the equilibrium state. The equilibrium current near the Fermi surface between two adjacent lattice points is
\begin{equation}
	\frac{\dif I_{rr'}}{\dif E}=-\frac{2e}{h}\sum_{i,j}\mathrm{Im}([t_{rr'}]_{ij}^*[G^r(E_F)-G^a(E_F)]_{ij}).
\end{equation}
For every single site $\vec{r}$, the current density can be obtained by summing all the $I_{rr'}$ up.

When differential voltage is applied, the equilibrium is broken, and the lesser Green's function is no longer described by the fluctuation-dissipation theorem but in the following form:
\begin{equation}
	G^<(E)=G^r(E)\Sigma^<(E)G^a(E).
\end{equation}
The total lesser self-energy $\Sigma^<(E)$ is contributed from all the leads, and the lesser self-energy in the $l$th lead is
\begin{equation}
	\Sigma^<_l(E)=\mi f_l(E)\Gamma_l.
\end{equation}
The current $I_{rr'}$ can thus be expressed as
\begin{equation}
	I_{rr'}=-\frac{2e}{\hbar}\int\frac{d E}{2\piup}\sum_{i,j}\mathrm{Re}([t_{rr'}]_{ij}^*[G^r\sum_{l\in\mathrm{all}}\mi f_l(E)\Gamma_lG^a]_{ij}).
\end{equation}
If the differential voltage raise the potential in the real contact L and decrease the potential in the real contact R, then the differential current in zero temperature is
\begin{equation}
	\frac{\dif I_{rr'}}{\dif V}=\frac{2e}{\hbar}\sum_{i,j}\mathrm{Im}([t_{rr'}]_{ij}^*[G^r(\frac{\Gamma_L-\Gamma_R}{2}+\sum_{i\in\mathrm{fic}}x_i\Gamma_i)G^a]_{ij}),
\end{equation}
where $x_i=\mu_i/V$.
\end{appendix}

\end{document}